\documentclass[aps,superscriptaddress,twocolumn,longbibliography]{revtex4-1}
\usepackage[pdftex]{graphicx}
\usepackage{color}
\graphicspath{{figures/}}
\newcommand{\VBG}{$V_{\mathrm{bg}}$~}
\newcommand{\VD}{$V_{\mathrm{d}}$~}
\newcommand{\VGone}{$V_{\mathrm{g1}}$~}
\newcommand{\VGtwo}{$V_{\mathrm{g2}}$~}
\newcommand{\CGone}{$C_{\mathrm{g1}}$~}
\newcommand{\CGtwo}{$C_{\mathrm{g2}}$~}
\newcommand{\CG}{$C_{\mathrm{g}}$~}
\newcommand{\LG}{$L_{\mathrm{g}}$~}
\newcommand{\SGG}{$S_{\mathrm{gg}}$~}
\newcommand{\IDS}{$I_{\mathrm{ds}}$~}
\newcommand{\VDIAG}{$V_{\mathrm{diag}}$~}
\newcommand{\TSi}{$T_{\mathrm{Si}}$~}

\begin{document}

\title{A hybrid metal/semiconductor electron pump \\ for quantum metrology}

\author{X. Jehl}
\email[]{xavier.jehl@cea.fr}
\author{B. Voisin}
\affiliation{SPSMS, UMR-E CEA / UJF-Grenoble 1, INAC, F-38054 Grenoble, France}
\author{T. Charron}
\affiliation{Laboratoire National d'Essais, F-78197 Trappes, France}
\author{P. Clapera}
\author{S. Ray}
\author{B. Roche}
\author{M. Sanquer}
\affiliation{SPSMS, UMR-E CEA / UJF-Grenoble 1, INAC, F-38054 Grenoble, France}
\author{S. Djordjevic}
\author{L. Devoille}
\affiliation{Laboratoire National d'Essais, F-78197 Trappes, France}
\author{R. Wacquez}
\author{M. Vinet}
\affiliation{CEA, LETI, Minatec campus, F-38054 Grenoble, France}

\begin{abstract}
Electron pumps capable of delivering a current higher than 100\,pA with sufficient accuracy are likely to become the direct mise en pratique of the possible new quantum definition of the ampere. Furthermore, they are essential for closing the quantum metrological triangle experiment which tests for possible corrections to the quantum relations linking $e$ and $h$, the electron charge and the Planck constant, to voltage, resistance and current. We present here single-island hybrid metal/semiconductor transistor pumps which combine the simplicity and efficiency of Coulomb blockade in metals with the unsurpassed performances of silicon switches. Robust and simple pumping at 650\,MHz and 0.5\,K is demonstrated. The pumped current obtained over a voltage bias range of 1.4\,mV corresponds to a relative deviation of 5$\times 10^{-4}$ from the calculated value, well within the 1.5$\times 10^{-3}$ uncertainty of the measurement setup. Multi-charge pumping can be performed. The simple design fully integrated in an industrial CMOS process makes it an ideal candidate for national measurement institutes to realize and share a future quantum ampere.
\end{abstract}
\maketitle

\section{INTRODUCTION}
Linking measurement units to quantum effects rather than artefacts is a major achievement of modern science\cite{Flowers2004}.
The electrical base unit of the S.I., the ampere, which is defined by the force acting between two ideal conductors, is not directly realized anymore. Instead, electrical metrology has been based since 1990 on voltage and resistance realized with the Josephson and quantum Hall effects~\cite{Taylor1989}. For the S.I. to fully benefit from their remarkable accuracy, a full set of quantum electrical units based on $e$ and the Planck constant $h$ could be adopted, where the ampere would be a number of elementary charges $e$ per unit time. Just like the speed of light was fixed when the meter was redefined, the fundamental constants $e$ and $h$ would then be fixed and the $mise$ $en$ $pratique$ of the new definition of the ampere would be made by using the combination of the Josephson and quantum Hall effects. A SET current standard based on an electron pump driven at a frequency $f$, producing a quantized current $I=Nef$ (where $N$ is an integer) with an amplitude of 100 pA or more could also be used for the realization of the new ampere. Furthermore, such a current is necessary for the quantum metrological triangle experiment which consists in achieving Ohm's law between the three effects used in quantum electrical metrology: Josephson effect (voltage), quantum Hall effect (resistance) and single electron tunneling effect (current). It aims to test the possible corrections to the relations linking $e$ and $h$ to the constants associated with three quantum effects: Josephson constant $K_{\text{J}}$, von Klitzing constant $R_{\text{K}}$ and an estimate of the electron charge~\cite{Piquemal2000,Keller2008}.

The understanding of the Coulomb blockade phenomenon~\cite{Grabert1992} rapidly led to the design of electron pumps and turnstiles both with fixed~\cite{Geerligs1990,Pothier1992} and tunable barriers~\cite{Kouwenhoven1991}. While turnstiles require a bias voltage to operate, pumps deliver a d.c. current even at zero bias. Metallic pumps with fixed height tunnel oxide barriers and multiple Coulomb islands reached a relative accuracy in the 10$^{-8}$ range, with up to 6 islands in series between 7 tunnel junctions~\cite{Keller1996,Keller1999}. Their complex operation was simplified by using only 3 junctions while lowering errors due to cotunneling with resistors~\cite{Lotkhov2001}. Despite these efforts the level of current remained below 10\,pA ($f\approx$60\,MHz), too small by at least one order of magnitude for a practical current source. They are also restricted to very low temperatures (below 50\,mK) and cannot accept a bias voltage without a severe degradation of their accuracy~\cite{Jehl2003}. With silicon technology, we previously demonstrated electron pumping in a device with two coupled islands and fixed barriers~\cite{Pierre2010}. Hybrid normal/superconducting turnstiles~\cite{Pekola2008} (which require a finite bias $V_d$) based on fixed barriers are also relatively slow but can be operated in parallel to increase the current~\cite{Maisi2009}. An alternative option using optically driven self-assembled quantum dots has also been demonstrated recently~\cite{Nevou2011}.
A different pumping scheme revived a strong interest in the field because it reached the desired level of current~\cite{Blumenthal2007,Fujiwara2008,Kaestner2008,Giblin2010} and recently an accuracy approaching the ppm level~\cite{Giblin2012}. It uses non-adiabatic effects in the population of a dynamical quantum dot~\cite{Kashcheyevs2010}. Operation in series of several such pumps have also been realized, with an integrated charge sensing scheme~\cite{Fricke2011}.
At the same time as metallic pumps with fixed barriers an experiment was performed which also uses orthodox Coulomb blockade but on a single semiconducting island connected by two tunable barriers~\cite{Kouwenhoven1991}. Using $\pi$ phase shifts on the two barriers, quantized currents were obtained, up to 20\,MHz. Although this pioneering work focuses on turnstile operation, electron pumping at zero bias was also briefly shown and discussed at the end of the report~\cite{Kouwenhoven1991}. Later on, a similar experiment with a silicon device~\cite{Ono2003} reached $f$=1\,MHz, and more recently the same idea was used but with $\pi$-shift only, resulting again in turnstile operation~\cite{Chan2011}. 

In this work we clarify this pumping mechanism with a novel type of electron pumps using industrial silicon technology. We show that the phase difference between the sinewaves applied to the two tunable barriers determine the number of electrons pumped per cycle and the direction of the current. A convincing evidence that this pumping mechanism is at play is the observation of an odd/even behaviour depending on the position of the working point with respect to the Coulomb peaks of the device.  
Since the pump studied here shows that it is possible to combine orthodox Coulomb blockade and high speed, it opens new opportunities for quantum metrology. 
This article is organized as follows. In section II, the samples technology and main characteristics are presented briefly, together with the results of d.c. electrical transport which unveil their electrostatic properties. Section III describes the pumping principle and the experimental results, and section IV focuses on experiments designed to assess the stability and accuracy of the pump.

\section{SAMPLE DESIGN AND d.c. TRANSPORT}
\subsection{Fully-depleted silicon-on-insulator technology}
\begin{figure}[!t]
\begin{center}
\includegraphics[width=\columnwidth, viewport=20 165 490 600,clip]{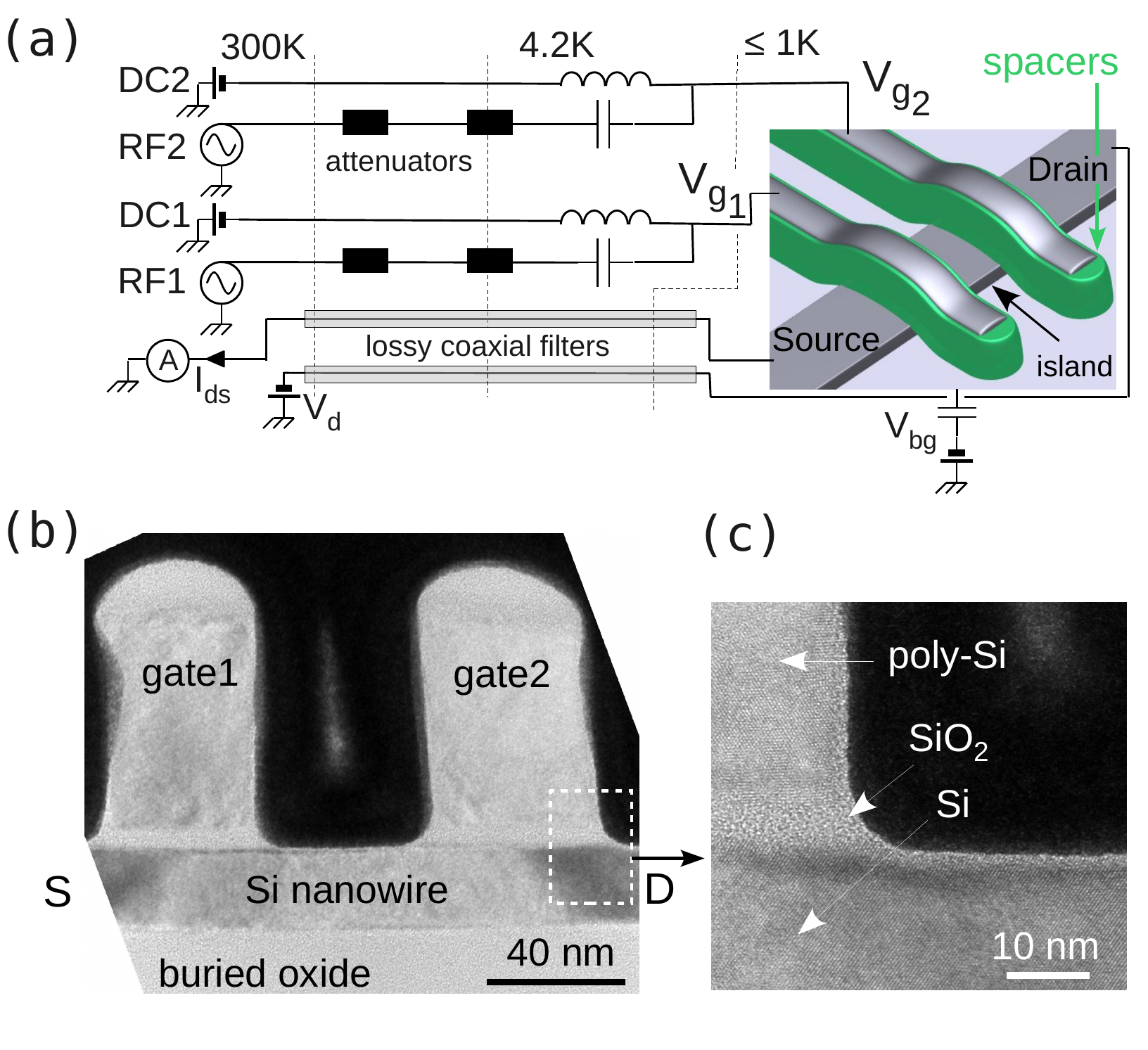}
\caption{(a) Schematic view of the wiring and sample layout. The single metallic island is controlled by two silicon transistors controlled by the gate voltages \VGone and \VGtwo on which dc and rf signals are applied. (b) and (c) Transmission electron micrographs of the two-gate structure after gate etching (spacers as well as silicidation are not shown here). The single-cristal silicon channel is isolated from the gates by 5\,nm of Si0$_2$. A backgate voltage \VBG is applied through the 145\,nm thick buried oxide of the silicon-on-insulator wafer.}
\label{fig1}
\end{center}
\end{figure}

A difficulty with quantized current sources comes from the necessity to have a system containing at some point a well defined number of charges $e$. Whether this relies on a high charging energy $E_c=e^2/C$ for metallic dots (here $C$ is the total capacitance of the island), or on the one-particle energy level spacing in a confinement potential created by gates, it requires the fabrication of a very small structure, at the limit of typical academic nanofabrication facilities. Silicon field-effect transistors (FETs) where gate length, channel width and thickness are all below 20\,nm are now developed and mass-produced by the microelectronics industry. We have fabricated our electron pumps with the advanced
Fully-Depleted Silicon-On-Insulator (FD-SOI) technology which just entered production for consumer electronics. Compared to conventional planar bulk technology, it offers a better electrostatic control over a very small channel, since the thin Si film forming the active area is isolated from the substrate by a buried oxide (BOX) layer of SiO$_2$, as shown in Fig.~\ref{fig1}b,c. This architecture is also very favorable for Coulomb blockade operation~\cite{Hofheinz2006a}. The very reduced volume of the transistor's channel yields a high charging energy, and it is well controlled by the gate electrode. The lever-arm parameter $\alpha$=\CG/$C$ is close to 1 at 300\,K (\CG is the gate capacitance). This translates into a slope $S$ for the sub-threshold current (an important figure of merit for transistors) approaching the ideal value set by thermal activation of the carriers: $S\approx ln(10)k_BT/e$=60\,mV/decade at $T$=300\,K ($k_B$ is the Boltzmann constant).     
In addition to the top gate acting as a tunable barrier (FET behaviour), biasing the substrate with a back-gate voltage \VBG allows to change both the threshold voltage and the tunnel coupling to source and drain by more than an order of magnitude. Thus the same transistor can be toggled between field-effect and single-electron behaviour~\cite{Roche2012}. The key steps of fabrication are nanowire lithography and etching, gate stack deposition (SiO$_2$ / N+ doped polysilicon) and patterning, silicon nitride spacer deposition and etch (highlighted in green in Fig.~\ref{fig1}a), source/drain epitaxy, doping and silicidation (NiSi). Then the standard back-end process includes embedding in SiO$_2$, opening and filling vias, metal lines (Cu) and finally Al pads.
To obtain a pump, two gates are designed in series with a small gap in between, as shown in Fig.~\ref{fig1}. During the silicidation process the gates and spacers act as a mask. As a result not only the source and drain become truly metallic, but also the island between the two gates. The critical dimensions of the devices measured in this work are given in Table~1.

\begin{table}
\begin{tabular}{|c||c|c|c|c|c|c|c|}
\hline name & $W$ & \TSi & \LG & \SGG & spacers & \CGone & \CGtwo \\ \hline
\hline s1 & 40 & 8 & 30 & 70 & 5+10 & 19.1 & 20.1 \\ 
\hline s2 & 60 & 20 & 50 & 50 & 5+10 & 17.2 & 15.5 \\
\hline s3 & 100 & 25 & 30 & 70 & 15 & 18.4 & 17.2 \\ 
\hline 
\end{tabular} 
\label{table}
\caption{Dimensions of the samples presented in this article, in nm. $W$ and \TSi stand for the nanowire's width and thickness, \LG is the gate length, \SGG the spacing between the gates, and "spacers" the spacers length. A single spacer was used for s3, while the standard industrial two-step spacer process was used for s1 and s2. \CGone and \CGtwo are the capacitive coupling in aF between the central island and respectively gates 1 and 2.}
\end{table}

\subsection{d.c. electrostatics and transport}
\begin{figure}[!t]
\begin{center}
\includegraphics[width=1.0\columnwidth, viewport=0 0 730 600,clip]{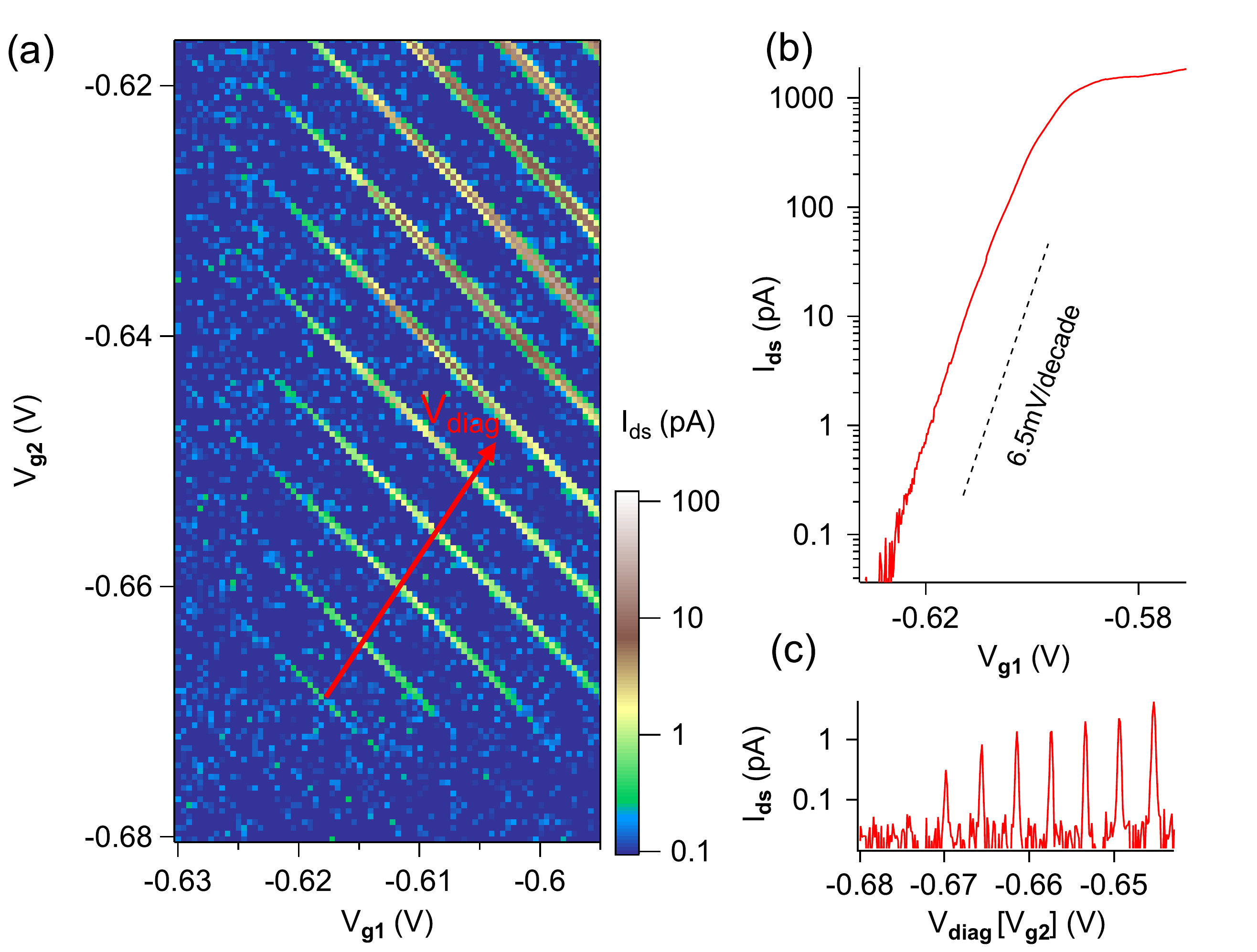}
\caption{Drain-source current \IDS for a small drain bias of 50\,$\mathrm{\mu}$V at 0.5\,K (sample s1). (a) 2D plot versus both gate voltages showing the pattern of segments due to the island capacitively coupled to both gates. The axis \VDIAG is defined by passing through the middle of the Coulomb segments. (b) \IDS versus \VGone for \VGtwo =0.8\,V. The steep sub-threshold slope over several decades illustrates the quality of the barrier below gate 1. (c) Cut of (a) along \VDIAG. The numerical values correspond to the projection on the \VGtwo axis.}
\label{fig2}
\end{center}
\end{figure}

The metallic island is by design capacitively coupled to the gates of both transistors, hence at low temperature the drain-source current \IDS versus gate voltages \VGone and \VGtwo is a series of antidiagonal lines, as illustrated in Fig. \ref{fig2}a at 0.5\,K. Each segment corresponds to the addition of an extra electron on the central island and is limited by the closure of the two FETs. In order to estimate the efficiency of each barrier separately we set one of them to +0.8\,V to minimize its contribution to the total current and sweep the other one. Such a measurement is shown in Fig. \ref{fig2}b. As expected for industrial FETs, we found a very steep current rise over more than 4 orders of magnitude. At low temperature the sub-threshold slope $S$ is no more set by the ideal thermal activation discussed before but saturates, mainly because the conduction band is broadened by disorder. Despite this, the recorded value of 6.5\,mV/decade at 0.5\,K is excellent and critical for electron pumping. Fig. \ref{fig2}c shows the periodicity of the current recorded when both gates are varied in order to follow the red diagonal line \VDIAG drawn on top of Fig. \ref{fig2}a. The values \CGone =19.1\,aF and \CGtwo =20.1\,aF extracted from this graph illustrate the symmetry of the coupling to both FET gates.   

\section{ELECTRON PUMPING}
\label{pompage}

\begin{figure}[!t]
\begin{center}
\includegraphics[width=\columnwidth, viewport=93 110 615 590,clip]{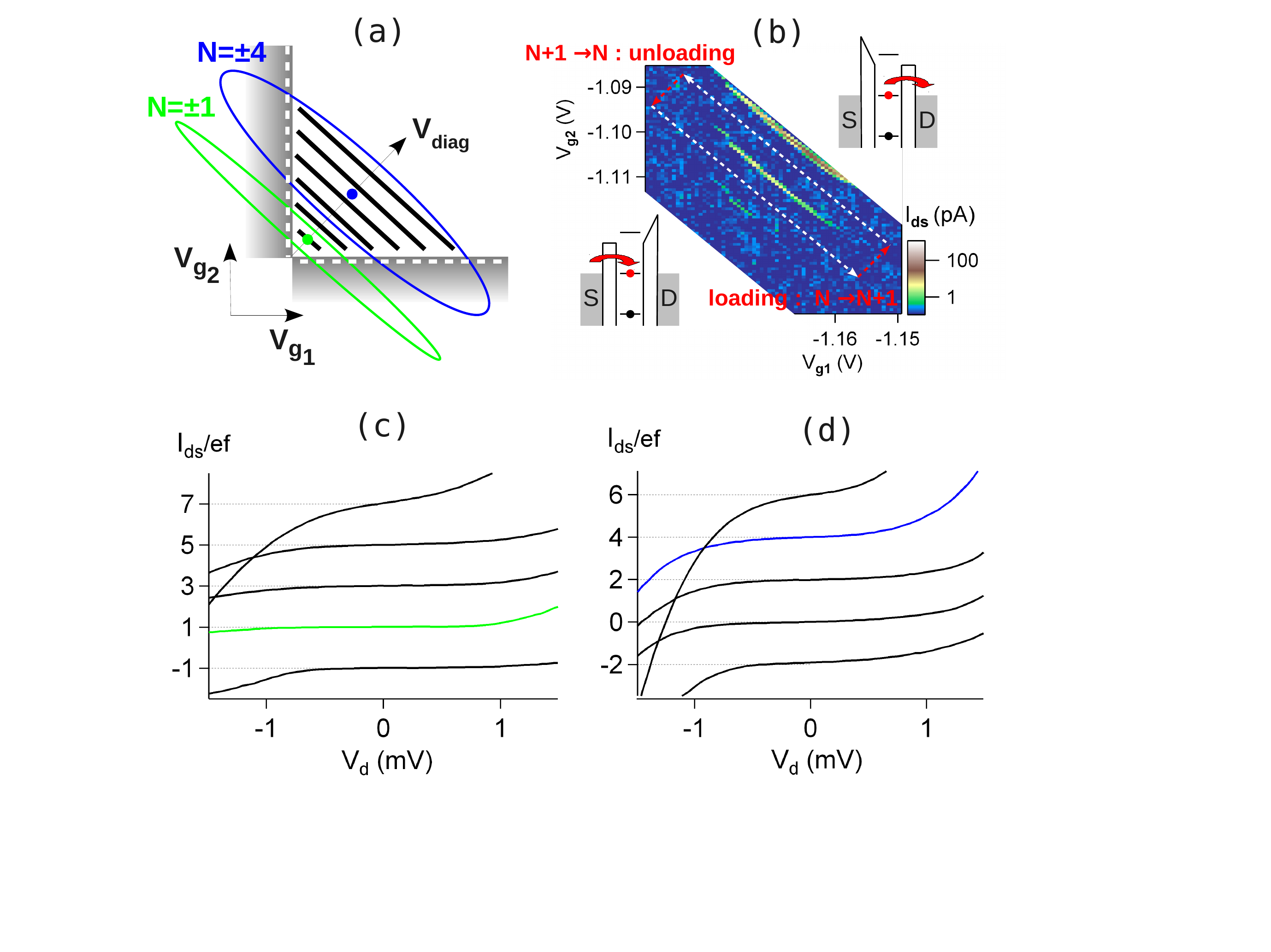}
\caption{(a) Schematics of pumping contours realized with phase-shifted RF signals on the two gates. Increasing the shift results in wider ellipses enclosing several Coulomb segments of the island (1 and 4 in the examples drawn in green and blue). The white dashed lines and shaded regions indicate the onset of conduction, respectively through gates 1 and 2. (b) \IDS versus \VGone and \VGtwo in sample s2 at 0.5K. The dashed lines indicate the 4 sequences of electron pumping. For the white lines the potential of the island is kept constant hence its population as well. The two red ones cross the Coulomb line and therefore require a population change $N<->N+1$. As sketched, these happen when one of the two barriers is very high, forcing electron transfer to occur through the other one. (c) Normalized current \IDS/$ef$ versus drain bias \VD for ellipses at $f$=54\,MHz centered near a Coulomb line, of equal amplitude on both gates (-26.5\,dBm) but different phase shifts. Plateaus at odd numbers of electrons are found at 1\,K since a growing but always odd number of lines is surrounded as the ellipses get wider. (d) Even plateaus obtained for ellipses centered in between two Coulomb lines and therefore enclosing an even number of lines.}
\label{fig3}
\end{center}
\end{figure}

\begin{figure*}[!t]
\begin{center}
\includegraphics[width=0.9\textwidth, viewport=0 00 850 600,clip]{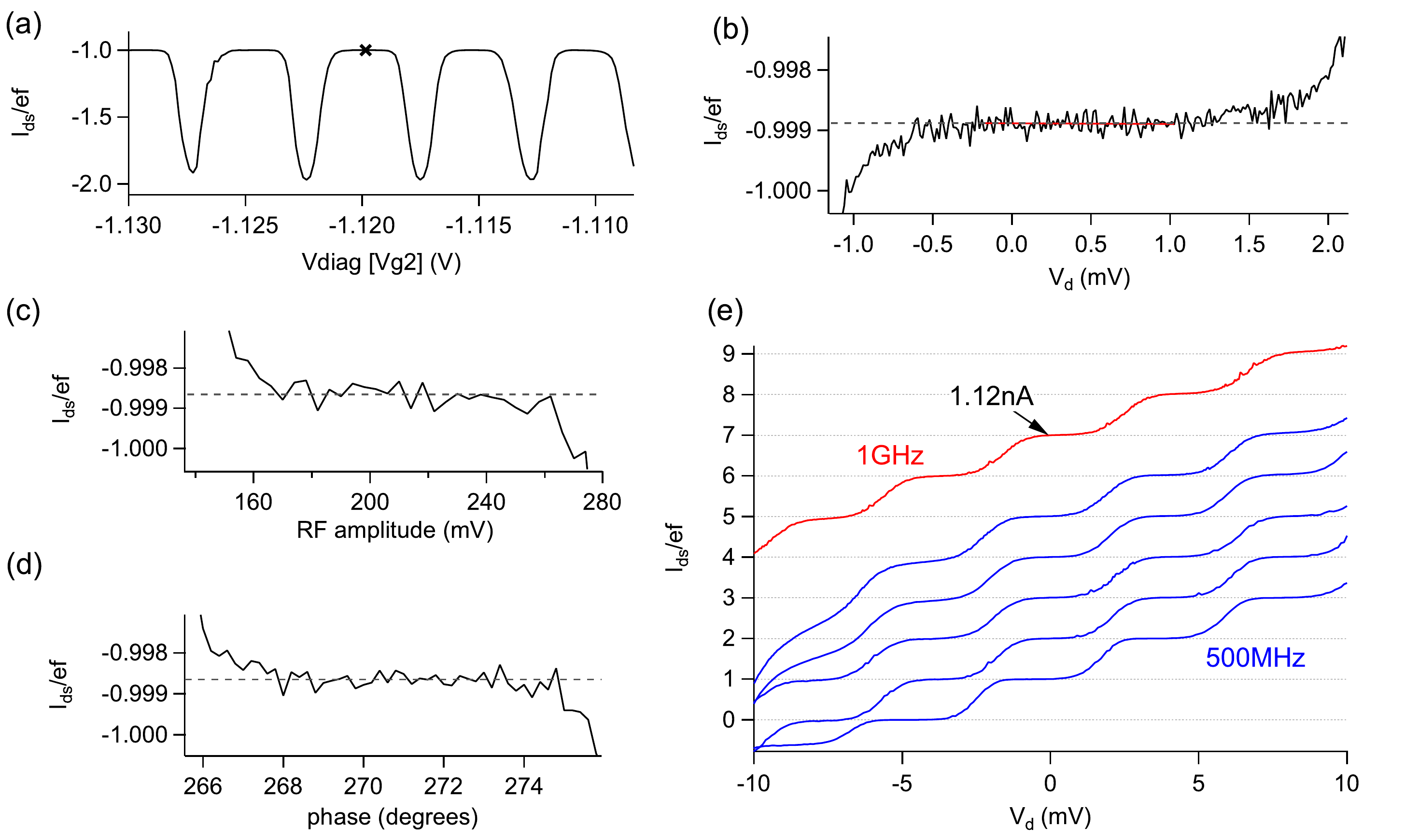}
\caption{(a) Pumped current \IDS$/ef$ versus \VDIAG for sample s2 and a dephasing of 189$^\circ$ chosen to pump N=-1 electron. The cross indicates the working point chosen for the measurements shown in (b) to (d), all performed also at 650\,MHz and 0.5\,K. (b) Pumped current versus bias \VD. The standard deviation of the linear fit across a region wider than 1\,mV (in red) sets a lower limit of 250\,G$\Omega$ to the resistance. (c) and (d) Pumped current versus rf amplitude on both gates and dephasing showing a wide range of stability for fixed N=-1 operation. For (b),(c) and (d) a dashed horizontal line is displayed as a guide to the eyes. (e) Same data as (b) but on sample s3 at 0.6\,K, up to 1\,GHz and at higher \VD. The Coulomb staircase is observed in the pumped current as an applied \VD corresponding to the charging energy of the island also forces the population to change.}
\label{fig4}
\end{center}
\end{figure*}

The principle of the pumping experiment is depicted in Fig. \ref{fig3}a,b. The white dashed line in Fig. \ref{fig3}a corresponds to the region where dc current measurement falls below the experimental noise floor because the barriers increase rapidly. Applying phase-shifted sinewaves on the gates corresponds to elliptic contours in the $V_{\mathrm{g1}}-V_{\mathrm{g2}}$ plane. They can be set to turn around a Coulomb peak without ever crossing a black segment where current flows. For this their center (set by the d.c. values of \VGone and \VGtwo) has to be chosen on or near the Coulomb segment and the extension has to be large enough, as shown by the ellipses in Fig. \ref{fig3}a. This contour has been used previously with silicon but with extremely opaque barriers (always larger than $10^{11}\Omega$), resulting in extremely slow (1\,MHz) and single-charge pumping\cite{Ono2003}. With barriers more than 6 orders of magnitude more transparent, we can pump faster and extend the contour to turn around several Coulomb lines, as illustrated by the blue ellipse in Fig. \ref{fig3}a. Instead of a net current $I=ef$ for the green ellipse, one expects $I=Nef$ where $N$ is the number of Coulomb lines enclosed by the contour ($N$=4 for the blue ellipse). The sign of the current is simply changed by turning in the other direction. Although never reported so far, the periodicity of the Coulomb oscillations should yield a different evolution of $N$ depending on the position of the contour center. With ellipses centered near a line, only odd numbers of lines are enclosed, while only even numbers occur for ellipses right in between two lines. These two situations have been experimentally realized. The corresponding pumped current plateaus are shown respectively in Figs. \ref{fig3}c,d. They follow exactly the expected odd/even behaviour.

The evolution of the slope of the plateaus as a function of different parameters have been investigated at high frequency. Results for sample s2 are shown in Fig. \ref{fig4}. The evolution of the pumped current versus \VDIAG is shown in Fig. \ref{fig4}a at 0.5\,K, for \VD =0. The rms amplitudes of the r.f. signals at 650\,MHz on the gates are 19\,mV and 19.8\,mV, in order to account for the small difference in couplings to the two gates, and therefore align the long axis of the ellipse with the Coulomb segments pattern. The phase difference between the a.c. gate voltages was set 189$^\circ$, in order to pump preferentialy -1 electron. When the ellipse encloses only one Coulomb segment, a quantized plateau at \IDS /$ef$=-1 is obtained. Moving along \VDIAG, there are regions where two segments are partially enclosed. In that case there is a tendency to pump more than one electron, hence the recorded current periodically increases. This period (approximately 4.1\,mV) is equal to the one extracted from the conductance measurements from which the gate capacitances given in Table\,1 were obtained. A working point indicated by the cross in Fig. \ref{fig4}a is chosen to study the dependence of the pumped current with bias \VD (Fig. \ref{fig4}b) and r.f. amplitude (Fig. \ref{fig4}c) and phase (Fig. \ref{fig4}d). An important figure of merit for a current source is its output resistance, corresponding to the flatness of the plateau in Fig. \ref{fig4}b. From the uncertainty in a linear fit of the data across a large span of \VD (1.2\,mV), in red, we obtain the lower limit of 250\,G$\mathrm{\Omega}$ for this resistance. The flatness of the plateaus versus r.f. amplitude and relative phase are important to check that the flat region is wide enough to be well within control. The data shown in Fig. \ref{fig4}c,d gives a comfortable working range of 70\,mV and 6$^\circ$.

Operation up to 1\,GHz and for several values of $N$ is shown in Fig. \ref{fig4}e at 0.6\,K. At \VD =0 a plateau is still obtained up to $N=7$, corresponding to a quantized current $I\approx$1.12\,nA. For this measurement with large $N$, i.e. a wide ellipse, the r.f. amplitude has to be increased significantly to maintain a large penetration into the regions shaded in grey in Fig. \ref{fig3}a, where the island population will eventually change. A maximum rms amplitude of 33.5\,mV was used (for $N$=7 at $f$=1\,GHz), which is possible at 600\,mK but could be challenging at significantly lower temperatures, where heating can occur\cite{Chan2011}. It is clear however that the flatness quickly degrades with increasing $N$, as a result of the inadequate trajectory looking too much like a circle. For better performance a custom trajectory performed with an arbitrary waveform generator could be used~\cite{Giblin2012}. 

While $N$ is determined near \VD =0 by the number of Coulomb lines enclosed by the ellipse, it can be changed with a sufficient \VD bias. This corresponds to the Coulomb staircase. In principle this is the case in any dot, however the barriers are often strongly affected by the bias and only the first diamond can be observed. In our system clear plateaus are still found up to 8\,mV, with a period of approximately 4\,mV (Fig. \ref{fig4}e) which can be directly interpreted as the charging energy (4\,meV), without any lever-arm parameter. This is in excellent agreement with Coulomb diamond measurements (not shown). This behaviour in \VD is characteristic of systems relying on orthodox Coulomb blockade, whether based on GaAs\cite{Kouwenhoven1991}, Si\cite{Chan2011} or a metal like here. It is strikingly different from the pumping scheme where the barriers are made more transparent than the quantum of conductance. In that case the middle island is fully connected to source or drain (hence there is no Coulomb blockade) and no \VD dependence is found, even up to 200\,mV\cite{Fujiwara2004}.
The possibility to observe clear plateaus with a bias voltage up to twice the charging energy is directly related to the robustness of the barriers in the OFF state, a figure of merit called drain-induced barrier lowering in microelectronics.

\section{STABILITY AND ACCURACY EXPERIMENTS}

\begin{figure}[!t]
\begin{center}
\includegraphics[width=\columnwidth, viewport=0 0 665 505,clip]{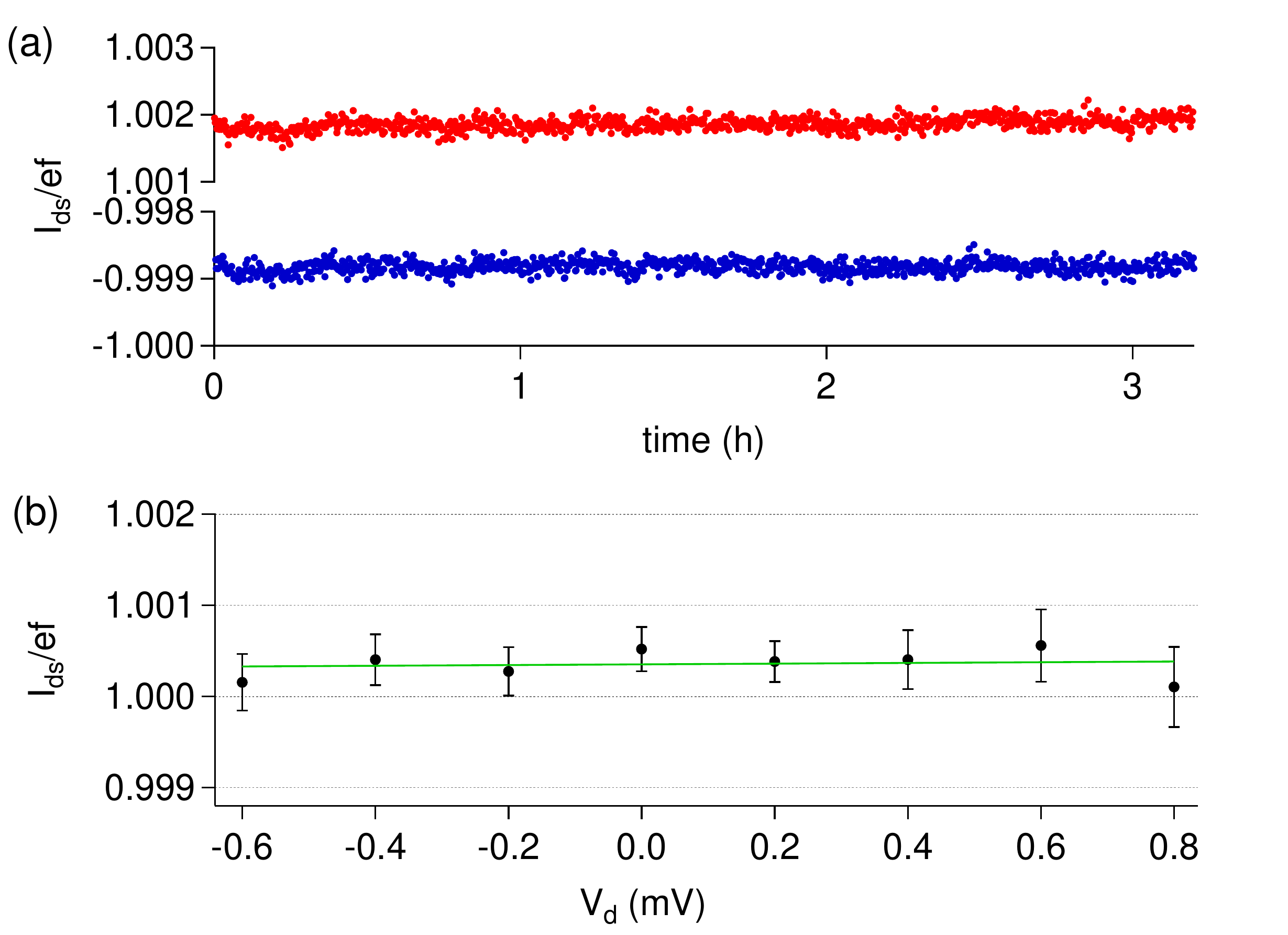}
\caption{(a) Alternating pumping experiment performed in sample s2 at 0.5\,K, 650\,MHz and for \VD=0\,mV. The dephasing is periodically set to 171$^\circ$ and 189$^\circ$, in order to pump respectively +1 (red) and -1 electron (blue) per cycle. This sequence is repeated over several hours. (b) Average values for + and -$ef$ pumping obtained from data similar to (a) but on 10 minutes only, and for several \VD. The error bars are twice the statistical standard deviation. This dataset yields a value of \IDS /$ef$=1.00036, well within the uncertainty of the measurement setup, and sets a lower limit of 6.8$\times10^{13}\,\mathrm{\Omega}$ for the resistance (standard deviation of the linear fit in green).}
\label{fig5}
\end{center}
\end{figure}

The accuracy of the pumped current has been assessed with alternating current techniques, in order to cancel the offset of the measurement chain.

Firstly, preliminary current measurements using a cryogenic current comparator (CCC) have been carried out on a fourth sample, at a reduced frequency of 50\, MHz with the CCC in internal feedback mode. Although this scheme does not allow an accurate measurement of the absolute value of the current, it allows to investigate the current steps flatness and the long-term stability of the device, shown in the supplementary material~\cite{Steck2008}. The Allan deviation calculated for a 14-hour long measurement exhibits a $\tau^{-1/2}$ variation with the averaging time $\tau$. This is a signature of a white noise regime over the whole range of measurement time, up to 14 hours (Figures 1 and 2 of the supplementary material). The statistical uncertainty can therefore be estimated by calculating the experimental deviation of the mean over the whole set of data. It results in a relative statistical uncertainty of 3 parts in 10$^6$ on the current value.

Secondly, calibrated commercial equipments have been used.
Fig. \ref{fig5}a shows the time evolution of the $\pm ef$ currents obtained by switching the phase alternatively between 171$^\circ$ and 189$^\circ$, at 0.5\,K, 650\,MHz and \VD =0. After changing the phase and waiting for 5 seconds, two points were recorded for each phase, with a 1\,s integration time.

In order to evaluate the robustness of the obtained value with bias voltage, this alternating technique is used but on the much shorter timescale of a few tens of minutes, for which averaging out two-by-two subsequent values for + and - $ef$ is beneficial (see supplementary material). Then this whole set of data points is averaged to yield a unique value for a certain \VD. This final dataset is shown in Fig. \ref{fig5}b. The standard deviation of a linear fit (in green in Fig. \ref{fig5}b) across the 1.4\,mV of \VD gives a lower limit of 6.8$\times10^{13}\,\mathrm{\Omega}$ for the resistance. Averaging these 7 values yields a final current \IDS /$ef$=1.00036 across a 1.4\,mV voltage bias span, at 650\,MHz and 0.5\,K, with a standard deviation of 0.00017. This value is well within the 0.15\% uncertainty in the gain of our measurement setup after calibration.

\section{CONCLUDING REMARKS} 
The first experiments presented here on an industrial CMOS electron pump have been performed mostly at 650\,MHz, in order to produce a pumped current of the order of 100\,pA, high enough to be relevant for the quantum metrological triangle experiment. It should be noted that the experiments were carried out without any radiation shielding around the sample inside the vacuum chamber of a dilution refrigerator. Although our best value falls well below the accuracy of the measurement system, this could significantly improve further measurements with a more refined setup. The temperature of 0.5\,K at which we did these experiments illustrates the small dimensions (high charging energy) of the nanowire transistors. However, a study in temperature will be needed now, in order to investigate the main cause of errors in the pumping mechanism. An important contribution to these errors may indeed come from the thermal broadening of the Coulomb lines. Another source of errors will arise from the finite steepness of the FETs sub-threshold slope. Clearly, when the working point is at an extremum of the ellipse, what matters is the ratio of the minimum resistance $R_{min}$ of one barrier and the maximum resistance $R_{max}$ reached by the other one~\cite{Ono2003}. For $R_{min}$ bound to be larger than $10^5\,\mathrm{\Omega}$ for Coulomb blockade operation, a resistance of $R_{max}=10^{12}\,\mathrm{\Omega}$ has to be reached for the other one if one targets a relative uncertainty of $10^{-7}$. This kind of performance is probably reachable at low temperature with the FDSOI technology which shows unrivaled performance in terms of leakage. A more quantitative study is necessary, using the simulation tools we have developed for our Coulomb blockade devices and single-charge impurities
~\cite{Pierre2009}. A realistic simulation of the electrostatic system and tunable FET barriers has been implemented recently and will be compared with the experimental data
The reproducibility and robustness of the fabrication technology ensures a consistent quality allowing very exhaustive experimental investigations, since these embedded samples do not age significantly. We have measured the same SET device based on SOI technology 6 years apart, and it has shown no significant modification in the pattern of Coulomb oscillations at low temperature. Another advantage of the nanowire transistors is that these pumps operate as FETs at 300\,K. The room temperature measurements provide valuable information on the samples (e.g. maximum conductance, symmetry of the barriers) which allows to sort and select samples for low temperature investigation among the thousands available.
A practical current standard based on this device can in addition benefit from straightforward integration with silicon technology. This includes on-chip CMOS circuits or drive electronics as well as r.f. design for 1-10\,GHz operation.
The authors acknowledge financial support from the French ANR under Project POESI, the EC FP7 FET-proactive NanoICT under Project AFSiD and the EMRP program under Project SIB-07 Quantum Ampere. The Nanosciences foundation is supporting this work with the PhD grant of P.C. We also thank Mark W Keller for his comments and careful reading of the manuscript.


\begin{thebibliography}{10}%
\makeatletter
\providecommand \@ifxundefined [1]{%
 \ifx #1\undefined \expandafter \@firstoftwo
 \else \expandafter \@secondoftwo
\fi
}%
\providecommand \@ifnum [1]{%
 \ifnum #1\expandafter \@firstoftwo
 \else \expandafter \@secondoftwo
\fi
}%
\providecommand \enquote [1]{``#1''}%
\providecommand \bibnamefont  [1]{#1}%
\providecommand \bibfnamefont [1]{#1}%
\providecommand \citenamefont [1]{#1}%
\providecommand\href[0]{\@sanitize\@href}%
\providecommand\@href[1]{\endgroup\@@startlink{#1}\endgroup\@@href}%
\providecommand\@@href[1]{#1\@@endlink}%
\providecommand \@sanitize [0]{\begingroup\catcode`\&12\catcode`\#12\relax}%
\@ifxundefined \pdfoutput {\@firstoftwo}{%
 \@ifnum{\z@=\pdfoutput}{\@firstoftwo}{\@secondoftwo}%
}{%
 \providecommand\@@startlink[1]{\leavevmode}%
 \providecommand\@@endlink[0]{}%
}{%
 \providecommand\@@startlink[1]{%
  \leavevmode
  \pdfstartlink
   attr{/Border[0 0 1 ]/H/I/C[0 1 1]}%
   user{/Subtype/Link/A<</Type/Action/S/URI/URI(#1)>>}%
  \relax
 }%
 \providecommand\@@endlink[0]{\pdfendlink}%
}%
\providecommand \url  [0]{\begingroup\@sanitize \@url }%
\providecommand \@url [1]{\endgroup\@href {#1}{\urlprefix}}%
\providecommand \urlprefix [0]{URL }%
\providecommand \Eprint[0]{\href }%
\@ifxundefined \urlstyle {%
  \providecommand \doi [1]{doi:\discretionary{}{}{}#1}%
}{%
  \providecommand \doi [0]{doi:\discretionary{}{}{}\begingroup
  \urlstyle{rm}\Url }%
}%
\providecommand \doibase [0]{http://dx.doi.org/}%
\providecommand \Doi[1]{\href{\doibase#1}}%
\providecommand \bibAnnote [3]{%
  \BibitemShut{#1}%
  \begin{quotation}\noindent
    \textsc{Key:}\ #2\\\textsc{Annotation:}\ #3%
  \end{quotation}%
}%
\providecommand \bibAnnoteFile [2]{%
  \IfFileExists{#2}{\bibAnnote {#1} {#2} {\input{#2}}}{}%
}%
\providecommand \typeout [0]{\immediate \write \m@ne }%
\providecommand \selectlanguage [0]{\@gobble}%
\providecommand \bibinfo [0]{\@secondoftwo}%
\providecommand \bibfield [0]{\@secondoftwo}%
\providecommand \translation [1]{[#1]}%
\providecommand \BibitemOpen[0]{}%
\providecommand \bibitemStop [0]{}%
\providecommand \bibitemNoStop [0]{.\EOS\space}%
\providecommand \EOS [0]{\spacefactor3000\relax}%
\providecommand \BibitemShut [1]{\csname bibitem#1\endcsname}%
\bibitem{Flowers2004}%
  \BibitemOpen
  \bibfield{author}{%
  \bibinfo {author} {\bibfnamefont{J.}\ \bibnamefont{Flowers}},\ }%
  \bibfield{title}{%
  \enquote{\bibinfo {title} {The route to atomic and quantum standards},}\ }%
  \bibfield{journal}{%
  \bibinfo {journal} {Science}\ }%
  \textbf{\bibinfo {volume} {306}},\ \bibinfo {pages} {1324} (\bibinfo
  {year} {2004})%
  \bibAnnoteFile{NoStop}{Flowers2004}%
\bibitem{Taylor1989}%
  \BibitemOpen
  \bibfield{author}{%
  \bibinfo {author} {\bibfnamefont{B.~N.}\ \bibnamefont{Taylor}}\ and\ \bibinfo
  {author} {\bibfnamefont{T.~J.}\ \bibnamefont{Witt}},\ }%
  \bibfield{title}{%
  \enquote{\bibinfo {title} {New international electrical reference standards
  based on the josephson and quantum hall effects},}\ }%
  \bibfield{journal}{%
  \bibinfo {journal} {Metrologia}\ }%
  \textbf{\bibinfo {volume} {26}},\ \bibinfo {pages} {47} (\bibinfo {year}
  {1989})%
  \bibAnnoteFile{NoStop}{Taylor1989}%
\bibitem{Piquemal2000}%
  \BibitemOpen
  \bibfield{author}{%
  \bibinfo {author} {\bibfnamefont{F.}~\bibnamefont{Piquemal}}\ and\ \bibinfo
  {author} {\bibfnamefont{G.}~\bibnamefont{Genev\`es}},\ }%
  \bibfield{title}{%
  \enquote{\bibinfo {title} {Argument for a direct realization of the quantum
  metrological triangle},}\ }%
  \bibfield{journal}{%
  \bibinfo {journal} {Metrologia}\ }%
  \textbf{\bibinfo {volume} {37}},\ \bibinfo {pages} {207} (\bibinfo {year}
  {2000})%
  \bibAnnoteFile{NoStop}{Piquemal2000}%
\bibitem{Keller2008}%
  \BibitemOpen
  \bibfield{author}{%
  \bibinfo {author} {\bibfnamefont{M.~W.}\ \bibnamefont{Keller}},\ }%
  \bibfield{title}{%
  \enquote{\bibinfo {title} {Current status of the quantum metrology
  triangle},}\ }%
  \bibfield{journal}{%
  \bibinfo {journal} {Metrologia}\ }%
  \textbf{\bibinfo {volume} {45}},\ \bibinfo {pages} {102} (\bibinfo {year}
  {2008})%
  \bibAnnoteFile{NoStop}{Keller2008}%
\bibitem{Grabert1992}%
  \BibitemOpen
  \bibfield{author}{%
  \bibinfo {author} {\bibfnamefont{H.}~\bibnamefont{Grabert}}\ and\ \bibinfo
  {author} {\bibfnamefont{M.~H.}\ \bibnamefont{Devoret}},\ }%
  \emph{\bibinfo {title} {{Single Charge Tunneling: Coulomb Blockade Phenomena
  in Nanostructures}}}\ (\bibinfo {publisher} {Plenum Press, New York},\
  \bibinfo {year} {1992})%
  \bibAnnoteFile{NoStop}{Grabert1992}%
\bibitem{Geerligs1990}%
  \BibitemOpen
  \bibfield{author}{%
  \bibinfo {author} {\bibfnamefont{L.~J.}\ \bibnamefont{Geerligs}}, \bibinfo
  {author} {\bibfnamefont{V.~F.}\ \bibnamefont{Anderegg}}, \bibinfo {author}
  {\bibfnamefont{P.~A.~M.}\ \bibnamefont{Holweg}}, \bibinfo {author}
  {\bibfnamefont{J.~E.}\ \bibnamefont{Mooij}}, \bibinfo {author}
  {\bibfnamefont{H.}~\bibnamefont{Pothier}}, \bibinfo {author}
  {\bibfnamefont{D.}~\bibnamefont{Esteve}}, \bibinfo {author}
  {\bibfnamefont{C.}~\bibnamefont{Urbina}},\ and\ \bibinfo {author}
  {\bibfnamefont{M.~H.}\ \bibnamefont{Devoret}},\ }%
  \bibfield{title}{%
  \enquote{\bibinfo {title} {Frequency-locked turnstile device for single
  electrons},}\ }%
  \bibfield{journal}{%
  \Doi{10.1103/PhysRevLett.64.2691}{\bibinfo {journal} {Phys. Rev. Lett.}}\ }%
  \textbf{\bibinfo {volume} {64}},\ \bibinfo {pages} {2691} (\bibinfo {year} {1990})%
  \bibAnnoteFile{NoStop}{Geerligs1990}%
\bibitem{Pothier1992}%
  \BibitemOpen
  \bibfield{author}{%
  \bibinfo {author} {\bibfnamefont{H.}~\bibnamefont{Pothier}}, \bibinfo
  {author} {\bibfnamefont{P.}~\bibnamefont{Lafarge}}, \bibinfo {author}
  {\bibfnamefont{C.}~\bibnamefont{Urbina}}, \bibinfo {author}
  {\bibfnamefont{D.}~\bibnamefont{Esteve}},\ and\ \bibinfo {author}
  {\bibfnamefont{M.~H.}\ \bibnamefont{Devoret}},\ }%
  \bibfield{title}{%
  \enquote{\bibinfo {title} {Single-electron pump based on charging effects},}\
  }%
  \bibfield{journal}{%
  \bibinfo {journal} {EPL (Europhysics Letters)}\ }%
  \textbf{\bibinfo {volume} {17}},\ \bibinfo {pages} {249} (\bibinfo {year}
  {1992})%
  \bibAnnoteFile{NoStop}{Pothier1992}%
\bibitem{Kouwenhoven1991}%
  \BibitemOpen
  \bibfield{author}{%
  \bibinfo {author} {\bibfnamefont{L.~P.}\ \bibnamefont{Kouwenhoven}}, \bibinfo
  {author} {\bibfnamefont{A.~T.}\ \bibnamefont{Johnson}}, \bibinfo {author}
  {\bibfnamefont{N.~C.}\ \bibnamefont{van~der Vaart}}, \bibinfo {author}
  {\bibfnamefont{C.~J. P.~M.}\ \bibnamefont{Harmans}},\ and\ \bibinfo {author}
  {\bibfnamefont{C.~T.}\ \bibnamefont{Foxon}},\ }%
  \bibfield{title}{%
  \enquote{\bibinfo {title} {Quantized current in a quantum-dot turnstile using
  oscillating tunnel barriers},}\ }%
  \bibfield{journal}{%
  \bibinfo {journal} {Phys. Rev. Lett.}\ }%
  \textbf{\bibinfo {volume} {67}},\ \bibinfo {pages} {1626} (\bibinfo
  {year} {1991})%
  \bibAnnoteFile{NoStop}{Kouwenhoven1991}%
\bibitem{Keller1996}%
  \BibitemOpen
  \bibfield{author}{%
  \bibinfo {author} {\bibfnamefont{M.~W.}\ \bibnamefont{Keller}}, \bibinfo
  {author} {\bibfnamefont{J.~M.}\ \bibnamefont{Martinis}}, \bibinfo {author}
  {\bibfnamefont{N.~M.}\ \bibnamefont{Zimmerman}},\ and\ \bibinfo {author}
  {\bibfnamefont{A.~H.}\ \bibnamefont{Steinbach}},\ }%
  \bibfield{title}{%
  \enquote{\bibinfo {title} {Accuracy of electron counting using a 7-junction
  electron pump},}\ }%
  \bibfield{journal}{%
  \bibinfo {journal} {Appl. Phys. Lett.}\ }%
  \textbf{\bibinfo {volume} {69}},\ \bibinfo {pages} {1804} (\bibinfo
  {year} {1996})%
  \bibAnnoteFile{NoStop}{Keller1996}%
\bibitem{Keller1999}%
  \BibitemOpen
  \bibfield{author}{%
  \bibinfo {author} {\bibfnamefont{M.~W.}\ \bibnamefont{Keller}}, \bibinfo
  {author} {\bibfnamefont{A.~L.}\ \bibnamefont{Eichenberger}}, \bibinfo
  {author} {\bibfnamefont{J.~M.}\ \bibnamefont{Martinis}},\ and\ \bibinfo
  {author} {\bibfnamefont{N.~M.}\ \bibnamefont{Zimmerman}},\ }%
  \bibfield{title}{%
  \enquote{\bibinfo {title} {A capacitance standard based on counting
  electrons},}\ }%
  \bibfield{journal}{%
  \bibinfo {journal} {Science}\ }%
  \textbf{\bibinfo {volume} {285}},\ \bibinfo {pages} {1706} (\bibinfo
  {year} {1999})%
  \bibAnnoteFile{NoStop}{Keller1999}%
\bibitem{Lotkhov2001}%
  \BibitemOpen
  \bibfield{author}{%
  \bibinfo {author} {\bibfnamefont{S.~V.}\ \bibnamefont{Lotkhov}}, \bibinfo
  {author} {\bibfnamefont{S.~A.}\ \bibnamefont{Bogoslovsky}}, \bibinfo {author}
  {\bibfnamefont{A.~B.}\ \bibnamefont{Zorin}},\ and\ \bibinfo {author}
  {\bibfnamefont{J.}~\bibnamefont{Niemeyer}},\ }%
  \bibfield{title}{%
  \enquote{\bibinfo {title} {Operation of a three-junction single-electron pump
  with on-chip resistors},}\ }%
  \bibfield{journal}{%
  \bibinfo {journal} {Appl. Phys. Lett.}\ }%
  \textbf{\bibinfo {volume} {78}},\ \bibinfo {pages} {946} (\bibinfo
  {year} {2001})%
  \bibAnnoteFile{NoStop}{Lotkhov2001}%
\bibitem{Jehl2003}%
  \BibitemOpen
  \bibfield{author}{%
  \bibinfo {author} {\bibfnamefont{X.}\ \bibnamefont{Jehl}}, \bibinfo
  {author} {\bibfnamefont{M.~W.}\ \bibnamefont{Keller}}, \bibinfo {author}
  {\bibfnamefont{R.~L.}\ \bibnamefont{Kautz}}, \bibinfo {author}
  {\bibfnamefont{J.}~\bibnamefont{Aumentado}},\ and\ \bibinfo {author}
  {\bibfnamefont{J.~M.}\ \bibnamefont{Martinis}},\ }%
  \bibfield{title}{%
  \enquote{\bibinfo {title} {Counting errors in a voltage-biased electron
  pump},}\ }%
  \bibfield{journal}{%
  \Doi{10.1103/PhysRevB.67.165331}{\bibinfo {journal} {Phys. Rev. B}}\ }%
  \textbf{\bibinfo {volume} {67}},\ \bibinfo {pages} {165331} ( \bibinfo {year} {2003})%
  \bibAnnoteFile{NoStop}{Jehl2003}%
\bibitem{Pierre2010}%
  \BibitemOpen
  \bibfield{author}{%
  \bibinfo {author} {\bibfnamefont{M.}~\bibnamefont{Pierre}}, \bibinfo {author}
  {\bibfnamefont{B.}~\bibnamefont{Roche}}, \bibinfo {author}
  {\bibfnamefont{X.}~\bibnamefont{Jehl}}, \bibinfo {author}
  {\bibfnamefont{R.}~\bibnamefont{Wacquez}}, \bibinfo {author}
  {\bibfnamefont{M.}~\bibnamefont{Sanquer}}, \bibinfo {author}
  {\bibfnamefont{M.}~\bibnamefont{Vinet}}, \bibinfo {author}
  {\bibfnamefont{N.}~\bibnamefont{Feltin}},\ and\ \bibinfo {author}
  {\bibfnamefont{L.}~\bibnamefont{Devoille}},\ }%
  \enquote{\bibinfo {title} {{Operation of a Silicon CMOS Electron Pump}},}\ p.\ \bibinfo {pages} {{755}},\ \bibinfo
  {note} {{2010 Conference on Precision Electromagnetic Measurements, Daejeon,
  South Korea, June 13-18, 2010}}%
  \bibAnnoteFile{NoStop}{Pierre2010}%
\bibitem{Pekola2008}%
  \BibitemOpen
  \bibfield{author}{%
  \bibinfo {author} {\bibfnamefont{J.~P.}\ \bibnamefont{Pekola}}, \bibinfo
  {author} {\bibfnamefont{J.~J.}\ \bibnamefont{Vartiainen}}, \bibinfo
  {author} {\bibfnamefont{M.}\ \bibnamefont{Mottonen}}, \bibinfo {author}
  {\bibfnamefont{O-P}\ \bibnamefont{Saira}}, \bibinfo {author}
  {\bibfnamefont{M.}\ \bibnamefont{Meschke}},\ and\ \bibinfo {author}
  {\bibfnamefont{D.~V.}\ \bibnamefont{Averin}},\ }%
  \bibfield{title}{%
  \enquote{\bibinfo {title} {Hybrid single-electron transistor as a source of
  quantized electric current},}\ }%
  \bibfield{journal}{%
  \bibinfo {journal} {Nat. Phys.}\ }%
  \textbf{\bibinfo {volume} {4}},\ \bibinfo {pages} {120--124} ( \bibinfo {year} {2008})%
  \bibAnnoteFile{NoStop}{Pekola2008}%
\bibitem{Maisi2009}%
  \BibitemOpen
  \bibfield{author}{%
  \bibinfo {author} {\bibfnamefont{V.}\ \bibnamefont{Maisi}}, \bibinfo
  {author} {\bibfnamefont{Y.~A.}\ \bibnamefont{Pashkin}}, \bibinfo {author}
  {\bibfnamefont{S.}\ \bibnamefont{Kafanov}}, \bibinfo {author}
  {\bibfnamefont{J.-S.}\ \bibnamefont{Tsai}},\ and\ \bibinfo {author}
  {\bibfnamefont{J.}\ \bibnamefont{Pekola}},\ }%
  \bibfield{title}{%
  \enquote{\bibinfo {title} {Parallel pumping of electrons},}\ }%
  \bibfield{journal}{%
  \bibinfo {journal} {New Journal of Physics}\ }%
  \textbf{\bibinfo {volume} {11}},\ \bibinfo {pages} {113057} (\bibinfo {year}
  {2009})%
  \bibAnnoteFile{NoStop}{Maisi2009}%
\bibitem{Nevou2011}%
  \BibitemOpen
  \bibfield{author}{%
  \bibinfo {author} {\bibfnamefont{L.}~\bibnamefont{Nevou}}, \bibinfo {author}
  {\bibfnamefont{V.}~\bibnamefont{Liverini}}, \bibinfo {author}
  {\bibfnamefont{P.}~\bibnamefont{Friedli}}, \bibinfo {author}
  {\bibfnamefont{F.}~\bibnamefont{Castellano}}, \bibinfo {author}
  {\bibfnamefont{A.}~\bibnamefont{Bismuto}}, \bibinfo {author}
  {\bibfnamefont{H.}~\bibnamefont{Sigg}}, \bibinfo {author}
  {\bibfnamefont{F.}~\bibnamefont{Gramm}}, \bibinfo {author}
  {\bibfnamefont{E.}~\bibnamefont{Muller}},\ and\ \bibinfo {author}
  {\bibfnamefont{J.}~\bibnamefont{Faist}},\ }%
  \bibfield{title}{%
  \enquote{\bibinfo {title} {Current quantization in an optically driven
  electron pump based on self-assembled quantum dots},}\ }%
  \bibfield{journal}{%
  \bibinfo {journal} {Nat. Phys.}\ }%
  \textbf{\bibinfo {volume} {7}},\ \bibinfo {pages} {423} ( \bibinfo {year} {2011})%
  \bibAnnoteFile{NoStop}{Nevou2011}%
\bibitem{Blumenthal2007}%
  \BibitemOpen
  \bibfield{author}{%
  \bibinfo {author} {\bibfnamefont{M.~D.}\ \bibnamefont{Blumenthal}}, \bibinfo
  {author} {\bibfnamefont{B.}~\bibnamefont{Kaestner}}, \bibinfo {author}
  {\bibfnamefont{L.}~\bibnamefont{Li}}, \bibinfo {author}
  {\bibfnamefont{S.}~\bibnamefont{Giblin}}, \bibinfo {author}
  {\bibfnamefont{T.~J. B.~M.}\ \bibnamefont{Janssen}}, \bibinfo {author}
  {\bibfnamefont{M.}~\bibnamefont{Pepper}}, \bibinfo {author}
  {\bibfnamefont{D.}~\bibnamefont{Anderson}}, \bibinfo {author}
  {\bibfnamefont{G.}~\bibnamefont{Jones}},\ and\ \bibinfo {author}
  {\bibfnamefont{D.~A.}\ \bibnamefont{Ritchie}},\ }%
  \bibfield{title}{%
  \enquote{\bibinfo {title} {Gigahertz quantized charge pumping},}\ }%
  \bibfield{journal}{%
  \bibinfo {journal} {Nat. Phys.}\ }%
  \textbf{\bibinfo {volume} {3}},\ \bibinfo {pages} {343} ( \bibinfo {year} {2007})%
  \bibAnnoteFile{NoStop}{Blumenthal2007}%
\bibitem{Fujiwara2008}%
  \BibitemOpen
  \bibfield{author}{%
  \bibinfo {author} {\bibfnamefont{A.}\ \bibnamefont{Fujiwara}}, \bibinfo
  {author} {\bibfnamefont{K.}\ \bibnamefont{Nishiguchi}},\ and\ \bibinfo
  {author} {\bibfnamefont{Y.}\ \bibnamefont{Ono}},\ }%
  \bibfield{title}{%
  \enquote{\bibinfo {title} {Nanoampere charge pump by single-electron ratchet
  using silicon nanowire metal-oxide-semiconductor field-effect transistor},}\
  }%
  \bibfield{journal}{%
  \Doi{10.1063/1.2837544}{\bibinfo {journal} {Appl. Phys. Lett.}}\ }%
  \textbf{\bibinfo {volume} {92}},\ \bibinfo {eid} {042102} (\bibinfo {year}
  {2008})%
  \bibAnnoteFile{NoStop}{Fujiwara2008}%
\bibitem{Kaestner2008}%
  \BibitemOpen
  \bibfield{author}{%
  \bibinfo {author} {\bibfnamefont{B.}~\bibnamefont{Kaestner}}, \bibinfo
  {author} {\bibfnamefont{V.}~\bibnamefont{Kashcheyevs}}, \bibinfo {author}
  {\bibfnamefont{S.}~\bibnamefont{Amakawa}}, \bibinfo {author}
  {\bibfnamefont{M.~D.}\ \bibnamefont{Blumenthal}}, \bibinfo {author}
  {\bibfnamefont{L.}~\bibnamefont{Li}}, \bibinfo {author} {\bibfnamefont{T.~J.
  B.~M.}\ \bibnamefont{Janssen}}, \bibinfo {author}
  {\bibfnamefont{G.}~\bibnamefont{Hein}}, \bibinfo {author}
  {\bibfnamefont{K.}~\bibnamefont{Pierz}}, \bibinfo {author}
  {\bibfnamefont{T.}~\bibnamefont{Weimann}}, \bibinfo {author}
  {\bibfnamefont{U.}~\bibnamefont{Siegner}},\ and\ \bibinfo {author}
  {\bibfnamefont{H.~W.}\ \bibnamefont{Schumacher}},\ }%
  \bibfield{title}{%
  \enquote{\bibinfo {title} {Single-parameter nonadiabatic quantized charge
  pumping},}\ }%
  \bibfield{journal}{%
  \Doi{10.1103/PhysRevB.77.153301}{\bibinfo {journal} {Phys. Rev. B}}\ }%
  \textbf{\bibinfo {volume} {77}},\ \bibinfo {pages} {153301} (\bibinfo {year} {2008})%
  \bibAnnoteFile{NoStop}{Kaestner2008}%
\bibitem{Giblin2010}%
  \BibitemOpen
  \bibfield{author}{%
  \bibinfo {author} {\bibfnamefont{S.~P.}\ \bibnamefont{Giblin}}, \bibinfo
  {author} {\bibfnamefont{S.~J.}\ \bibnamefont{Wright}}, \bibinfo {author}
  {\bibfnamefont{J.~D.}\ \bibnamefont{Fletcher}}, \bibinfo {author}
  {\bibfnamefont{M.}~\bibnamefont{Kataoka}}, \bibinfo {author}
  {\bibfnamefont{M.}~\bibnamefont{Pepper}}, \bibinfo {author} {\bibfnamefont{T.~J.
  B.~M.}\ \bibnamefont{Janssen}}, \bibinfo {author} {\bibfnamefont{D.~A.}\
  \bibnamefont{Ritchie}}, \bibinfo {author} {\bibfnamefont{C.~A.}\
  \bibnamefont{Nicoll}}, \bibinfo {author}
  {\bibfnamefont{D.}~\bibnamefont{Anderson}},\ and\ \bibinfo {author}
  {\bibfnamefont{G.~A.~C.}\ \bibnamefont{Jones}},\ }%
  \bibfield{title}{%
  \enquote{\bibinfo {title} {An accurate high-speed single-electron quantum dot
  pump},}\ }%
  \bibfield{journal}{%
  \bibinfo {journal} {New Journal of Physics}\ }%
  \textbf{\bibinfo {volume} {12}},\ \bibinfo {pages} {073013} (\bibinfo {year}
  {2010})%
  \bibAnnoteFile{NoStop}{Giblin2010}%
\bibitem{Giblin2012}%
  \BibitemOpen
  \bibfield{author}{%
  \bibinfo {author} {\bibfnamefont{S.P.}\ \bibnamefont{Giblin}}, \bibinfo
  {author} {\bibfnamefont{M.}~\bibnamefont{Kataoka}}, \bibinfo {author}
  {\bibfnamefont{J.D.}\ \bibnamefont{Fletcher}}, \bibinfo {author}
  {\bibfnamefont{P.}~\bibnamefont{See}}, \bibinfo {author}
  {\bibfnamefont{T.J.B.M.}\ \bibnamefont{Janssen}}, \bibinfo {author}
  {\bibfnamefont{J.P.}\ \bibnamefont{Griffiths}}, \bibinfo {author}
  {\bibfnamefont{G.A.C.}\ \bibnamefont{Jones}}, \bibinfo {author}
  {\bibfnamefont{I.}~\bibnamefont{Farrer}},\ and\ \bibinfo {author}
  {\bibfnamefont{D.A.}\ \bibnamefont{Ritchie}},\ }%
  \bibfield{title}{%
  \enquote{\bibinfo {title} {Towards a quantum representation of the ampere
  using single electron pumps},}\ }%
  \bibfield{journal}{%
  \bibinfo {journal} {Nat. Commun.}\ }%
  \textbf{\bibinfo {volume} {3}},\ \bibinfo {pages} {930} ( \bibinfo {year} {2012})%
  \bibAnnoteFile{NoStop}{Giblin2012}%
\bibitem{Kashcheyevs2010}%
  \BibitemOpen
  \bibfield{author}{%
  \bibinfo {author} {\bibfnamefont{V.}\ \bibnamefont{Kashcheyevs}}\
  and\ \bibinfo {author} {\bibfnamefont{B.}\ \bibnamefont{Kaestner}},\ }%
  \bibfield{title}{%
  \enquote{\bibinfo {title} {Universal decay cascade model for dynamic quantum
  dot initialization},}\ }%
  \bibfield{journal}{%
  \Doi{10.1103/PhysRevLett.104.186805}{\bibinfo {journal} {Phys. Rev. Lett.}}\
  }%
  \textbf{\bibinfo {volume} {104}},\ \bibinfo {pages} {186805} ( \bibinfo {year} {2010})%
  \bibAnnoteFile{NoStop}{Kashcheyevs2010}%
\bibitem{Fricke2011}%
  \BibitemOpen
  \bibfield{author}{%
  \bibinfo {author} {\bibfnamefont{L.}\ \bibnamefont{Fricke}}, \bibinfo
  {author} {\bibfnamefont{F.}\ \bibnamefont{Hohls}}, \bibinfo {author}
  {\bibfnamefont{N.}\ \bibnamefont{Ubbelohde}}, \bibinfo {author}
  {\bibfnamefont{B.}\ \bibnamefont{Kaestner}}, \bibinfo {author}
  {\bibfnamefont{V.}\ \bibnamefont{Kashcheyevs}}, \bibinfo {author}
  {\bibfnamefont{C.}\ \bibnamefont{Leicht}}, \bibinfo {author}
  {\bibfnamefont{P.}\ \bibnamefont{Mirovsky}}, \bibinfo {author}
  {\bibfnamefont{K.}\ \bibnamefont{Pierz}}, \bibinfo {author}
  {\bibfnamefont{H.~W.}\ \bibnamefont{Schumacher}},\ and\ \bibinfo {author}
  {\bibfnamefont{R.~J.}\ \bibnamefont{Haug}},\ }%
  \bibfield{title}{%
  \enquote{\bibinfo {title} {Quantized current source with mesoscopic
  feedback},}\ }%
  \bibfield{journal}{%
  \Doi{10.1103/PhysRevB.83.193306}{\bibinfo {journal} {Phys. Rev. B}}\ }%
  \textbf{\bibinfo {volume} {83}},\ \bibinfo {pages} {193306} ( \bibinfo {year} {2011})%
  \bibAnnoteFile{NoStop}{Fricke2011}%
\bibitem{Ono2003}%
  \BibitemOpen
  \bibfield{author}{%
  \bibinfo {author} {\bibfnamefont{Y.}\ \bibnamefont{Ono}}\ and\ \bibinfo
  {author} {\bibfnamefont{Y.}\ \bibnamefont{Takahashi}},\ }%
  \bibfield{title}{%
  \enquote{\bibinfo {title} {Electron pump by a combined
  single-electron/field-effect- transistor structure},}\ }%
  \bibfield{journal}{%
  \Doi{10.1063/1.1556558}{\bibinfo {journal} {Appl. Phys. Lett.}}\ }%
  \textbf{\bibinfo {volume} {82}},\ \bibinfo {pages} {1221--1223} (\bibinfo
  {year} {2003})%
  \bibAnnoteFile{NoStop}{Ono2003}%
\bibitem{Chan2011}%
  \BibitemOpen
  \bibfield{author}{%
  \bibinfo {author} {\bibfnamefont{K.~W.}\ \bibnamefont{Chan}}, \bibinfo
  {author} {\bibfnamefont{M.}~\bibnamefont{M\"{o}tt\"{o}nen}}, \bibinfo
  {author} {\bibfnamefont{A.}~\bibnamefont{Kemppinen}}, \bibinfo {author}
  {\bibfnamefont{N.~S.}\ \bibnamefont{Lai}}, \bibinfo {author}
  {\bibfnamefont{K.~Y.}\ \bibnamefont{Tan}}, \bibinfo {author}
  {\bibfnamefont{W.~H.}\ \bibnamefont{Lim}},\ and\ \bibinfo {author}
  {\bibfnamefont{A.~S.}\ \bibnamefont{Dzurak}},\ }%
  \bibfield{title}{%
  \enquote{\bibinfo {title} {Single-electron shuttle based on a silicon quantum
  dot},}\ }%
  \bibfield{journal}{%
  \bibinfo {journal} {Appl. Phys. Lett.}\ }%
  \textbf{\bibinfo {volume} {98}},\ \bibinfo {eid} {212103} (\bibinfo {year}
  {2011})%
  \bibAnnoteFile{NoStop}{Chan2011}%
\bibitem{Hofheinz2006a}%
  \BibitemOpen
  \bibfield{author}{%
  \bibinfo {author} {\bibfnamefont{M.}~\bibnamefont{Hofheinz}}, \bibinfo
  {author} {\bibfnamefont{X.}~\bibnamefont{Jehl}}, \bibinfo {author}
  {\bibfnamefont{M.}~\bibnamefont{Sanquer}}, \bibinfo {author}
  {\bibfnamefont{G.}~\bibnamefont{Molas}}, \bibinfo {author}
  {\bibfnamefont{M.}~\bibnamefont{Vinet}},\ and\ \bibinfo {author}
  {\bibfnamefont{S.}~\bibnamefont{Deleonibus}},\ }%
  \bibfield{title}{%
  \enquote{\bibinfo {title} {Simple and controlled single electron transistor
  based on doping modulation in silicon nanowires},}\ }%
  \bibfield{journal}{%
  \Doi{10.1063/1.2358812}{\bibinfo {journal} {Appl. Phys. Lett.}}\ }%
  \textbf{\bibinfo {volume} {89}},\ \bibinfo {eid} {143504} (\bibinfo {year}
  {2006})%
  \bibAnnoteFile{NoStop}{Hofheinz2006a}%
\bibitem{Roche2012}%
  \BibitemOpen
  \bibfield{author}{%
  \bibinfo {author} {\bibfnamefont{B.}~\bibnamefont{Roche}}, \bibinfo {author}
  {\bibfnamefont{B.}~\bibnamefont{Voisin}}, \bibinfo {author}
  {\bibfnamefont{X.}~\bibnamefont{Jehl}}, \bibinfo {author}
  {\bibfnamefont{R.}~\bibnamefont{Wacquez}}, \bibinfo {author}
  {\bibfnamefont{M.}~\bibnamefont{Sanquer}}, \bibinfo {author}
  {\bibfnamefont{M.}~\bibnamefont{Vinet}}, \bibinfo {author}
  {\bibfnamefont{V.}~\bibnamefont{Deshpande}},\ and\ \bibinfo {author}
  {\bibfnamefont{B.}~\bibnamefont{Previtali}},\ }%
  \bibfield{title}{%
  \enquote{\bibinfo {title} {A tunable, dual mode field-effect or single
  electron transistor},}\ }%
  \bibfield{journal}{%
  \bibinfo {journal} {Appl. Phys. Lett.}\ }%
  \textbf{\bibinfo {volume} {100}},\ \bibinfo {pages} {032107} (\bibinfo {year}
  {2012})%
  \bibAnnoteFile{NoStop}{Roche2012}%
\bibitem{Fujiwara2004}%
  \BibitemOpen
  \bibfield{author}{%
  \bibinfo {author} {\bibfnamefont{A.}\ \bibnamefont{Fujiwara}}, \bibinfo
  {author} {\bibfnamefont{N.}\ \bibnamefont{Zimmerman}}, \bibinfo {author}
  {\bibfnamefont{Y.}\ \bibnamefont{Ono}},\ and\ \bibinfo {author}
  {\bibfnamefont{Y.}\ \bibnamefont{Takahashi}},\ }%
  \bibfield{title}{%
  \enquote{\bibinfo {title} {Current quantization due to single-electron
  transfer in si-wire charge-coupled devices},}\ }%
  \bibfield{journal}{%
  \Doi{10.1063/1.1650036}{\bibinfo {journal} {Appl. Phys. Lett.}}\ }%
  \textbf{\bibinfo {volume} {84}},\ \bibinfo {pages} {1323} (\bibinfo
  {year} {2004})%
  \bibAnnoteFile{NoStop}{Fujiwara2004}%
\bibitem{Steck2008}%
  \BibitemOpen
  \bibfield{author}{%
  \bibinfo {author} {\bibfnamefont{B.}~\bibnamefont{Steck}}, \bibinfo {author}
  {\bibfnamefont{A.}~\bibnamefont{Gonzalez-Cano}}, \bibinfo {author}
  {\bibfnamefont{N.}~\bibnamefont{Feltin}}, \bibinfo {author}
  {\bibfnamefont{L.}~\bibnamefont{Devoille}}, \bibinfo {author}
  {\bibfnamefont{F.}~\bibnamefont{Piquemal}}, \bibinfo {author}
  {\bibfnamefont{S.}~\bibnamefont{Lotkhov}},\ and\ \bibinfo {author}
  {\bibfnamefont{A.~B.}\ \bibnamefont{Zorin}},\ }%
  \bibfield{title}{%
  \enquote{\bibinfo {title} {Characterization and metrological investigation of
  an r-pump with driving frequencies up to 100 mhz},}\ }%
  \bibfield{journal}{%
  \bibinfo {journal} {Metrologia}\ }%
  \textbf{\bibinfo {volume} {45}},\ \bibinfo {pages} {482} (\bibinfo {year}
  {2008})%
  \bibAnnoteFile{NoStop}{Steck2008}%
\bibitem{Pierre2009}%
  \BibitemOpen
  \bibfield{author}{%
  \bibinfo {author} {\bibfnamefont{M.}~\bibnamefont{Pierre}}, \bibinfo {author}
  {\bibfnamefont{M.}~\bibnamefont{Hofheinz}}, \bibinfo {author}
  {\bibfnamefont{X.}~\bibnamefont{Jehl}}, \bibinfo {author}
  {\bibfnamefont{M.}~\bibnamefont{Sanquer}}, \bibinfo {author}
  {\bibfnamefont{G.}~\bibnamefont{Molas}}, \bibinfo {author}
  {\bibfnamefont{M.}~\bibnamefont{Vinet}},\ and\ \bibinfo {author}
  {\bibfnamefont{S.}~\bibnamefont{Deleonibus}},\ }%
  \bibfield{title}{%
  \enquote{\bibinfo {title} {Background charges and quantum effects in quantum
  dots transport spectroscopy},}\ }%
  \bibfield{journal}{%
  \Doi{10.1140/epjb/e2009-00258-4}{\bibinfo {journal} {European Physical
  Journal B}}\ }%
  \textbf{\bibinfo {volume} {70}},\ \bibinfo {pages} {475} ( \bibinfo {year} {2009})%
  \bibAnnoteFile{NoStop}{Pierre2009}%
\end{thebibliography}

%

\end{document}